\documentstyle[sprocl]{article}

\def\NPB{{\em Nucl. Phys.} B}
\def\PLB{{\em Phys. Lett.}  B}
\def\PRL{\em Phys. Rev. Lett.}
\def\PRD{{\em Phys. Rev.} D}

\def\be{\begin{equation}}
\def\ee{\end{equation}}
\def\bea{\begin{eqnarray}}
\def\eea{\end{eqnarray}}

\def\la{\hbox{{\lower -2.5pt\hbox{$<$}}\hskip -8pt\raise
-2.5pt\hbox{$\sim$}}}
\def\ga{\hbox{{\lower -2.5pt\hbox{$>$}}\hskip -8pt\raise
-2.5pt\hbox{$\sim$}}}

\begin{document}

\title{The Mystery of Ultra-High Energy Cosmic Rays}

\author{A. V. OLINTO}

\address{Department of Astronomy \& Astrophysics, \\ \& Enrico Fermi
Institute, \\
The University of Chicago, Chicago, IL 60637, USA\\E-mail:
olinto@oddjob.uchicago.edu}

\maketitle\abstracts{ 
The origin of cosmic rays with energies higher than 10$^{20}$ eV
remains a mystery. Accelerating particles up to these
energies is a challenge even for the most energetic
astrophysical objects known. While the isotropy in 
arrival directions argues for an extra-galactic origin, the
photon-pion production off the cosmic background radiation
limits the sources of such particles to systems less than
50 Mpc away from us. The combination of large gyroradii,
efficient energy losses, and isotropic arrival directions
defies most of the proposed astrophysical accelerators as
well as the more exotic alternatives. I briefly  review 
theoretical models for
the acceleration and propagation of ultra-high-energy
cosmic-rays and discuss the potential for future
observatories to resolve this
cosmic mystery.
}

\section{Introduction}
The origin of cosmic rays with energies above $10^{20}$ eV is an
intriguing mystery. At present, about 20
events above $10^{20}$ eV  have been reported worldwide  by  
experiments such as the High Resolution Fly's Eye, AGASA, Fly's Eye, 
Haverah Park, Yakutsk, and Volcano Ranch.  (For recent reviews of these
observations see, e.g., [1]).  The
unexpected flux above
$\sim 7 \times 10^{19}$ eV \cite{takeda98}  shows no sign of the
Greisen-Zatsepin-Kuzmin (GZK) cutoff.\cite{GZK66}  A cutoff should be
present if these  ultra-high energy particles are protons, nuclei, or
photons from  extragalactic sources.  Cosmic ray protons of energies
above  a few
$10^{19}$ eV lose energy to photopion production off the cosmic microwave
background (CMB) and cannot originate further than about $50\,$Mpc away
from Earth. Nuclei  are photodisintegrated on shorter distances due to
the infrared background while the radio background
constrains photons to originate from even closer systems.

In addition to the presence of events past the GZK cutoff, the 
arrival directions of the highest energy events show no clear angular
correlation with any of the plausible optical counterparts  such as
sources in the Galactic plane, the Local Group, or the Local
Supercluster.  If these events are protons, their arrival direction should
point back to their sources, but unlike luminous structures in a 50 Mpc
radius around us, the distribution of the highest energy events is
isotropic.

At these high energies the Galactic and extragalactic magnetic fields
should not affect the orbits significantly. Protons at
$10^{20}$ eV propagate mainly in straight lines as they traverse the
Galaxy since their gyroradii are $\sim $ 100 kpc in $ \mu$G  fields which
is typical in the Galactic disk  so they should
point back to their sources within a few degrees.  Extragalactic fields
are expected to be $\ll \mu$G, and induce at
most  $\sim$ 1$^o$ deviation from the source. Even if
the Local Supercluster has relatively strong fields, the highest energy
events may deviate at most $\sim$ 10$^o$.\cite{RKB98}  

If astrophysical sources cannot explain these observations, the
exciting alternative involves physics beyond the standard model of
particle physics. Not only the origin of these  particles  may be due to
physics beyond the standard model, but their
existence  can be used to constrain extensions of the standard model such
as violations of Lorentz invariance. 

The absence of a GZK cutoff and the isotropy of arrival directions are
some of the challenges that models for the origin of UHECRs face.
This mystery has generated a number of proposals but no model can
claim victory at this point. The exact shape of the spectrum at the
highest energies as well as a clear composition determination awaits
future observatories such as the Pierre Auger Project and
the proposed satellites OWL and Airwatch.

In this talk, I briefly review the models that attempt to solve this
mystery. For more extensive reviews, see [5].

\section{Astrophysical Zevatrons}

These challenging observations have generated two different
proposals to reaching a solution:  A
bottom-up approach involves looking for {\it Zevatrons}, possible
acceleration sites in known astrophysical objects that can reach ZeV
energies,  while a top-down approach involves the decay of very high mass
relics from the early universe and physics beyond the standard model of
particle physics.

Cosmic rays can be accelerated in
astrophysical plasmas when large-scale macroscopic motions, such
as shocks and turbulent flows, are transferred to individual particles.
The maximum energy of accelerated  particles,
$E_{\rm max}$, can be estimated by requiring that the gyroradius of the
particle be contained in the acceleration region: $E_{\rm max} = Ze \, B
\, L$, where  $Ze$ is the charge of the particle, $B$ is the strength  and
$L$ the  coherence length of the magnetic field embedded in the plasma.
For $E_{max} \ga 10^{20}$ eV and $Z \sim 1$, the only known
astrophysical sources with reasonable  $B L $ products   are neutron
stars ($B \sim 10^{13}$ G, $L \sim 10$ km),  active galactic nuclei (AGNs)
($B \sim 10^{4}$ G, $L \sim 10$ AU), radio lobes of AGNs ($B \sim
0.1\mu$G, $L \sim 10$ kpc), and clusters of galaxies ($B \sim \mu$G, $L
\sim 100$ kpc). 

{\it Clusters of Galaxies:}
Cluster shocks are
reasonable sites to consider for ultra-high energy cosmic ray (UHECR)
acceleration, since  particles with energy up to $E_{max}$ can be
contained by cluster fields. However, efficient losses due to 
photopion production off the CMB during the propagation inside the
cluster limit UHECRs in cluster shocks  to reach at most
$\sim$ 10 EeV.\cite{KRJ96KRB97}

{\it AGN Radio Lobes:}
Next on the list of plausible Zevatrons
are extremely powerful radio galaxies  \cite{BS87B97}. 
Jets from the central black-hole of an active galaxy end at a termination
shock where the interaction of the jet with the intergalactic medium
forms radio lobes and  `hot spots'. Of special interest are the most
powerful AGNs where shocks can accelerate particles to energies well
above an EeV via the first-order Fermi mechanism. These sources may be
responsible for the flux of UHECRs up to the GZK cutoff.\cite{RB93} 

A nearby specially powerful source may be able to reach energies  past the
cutoff.  However, extremely powerful AGNs with radio lobes
and hot spots are rare and far apart. The closest known object is M87 in
the Virgo cluster ($\sim$ 18 Mpc away)  and could be a main source of
UHECRs.  Although a single nearby source can fit the  spectrum 
for a given strength and structure of the intergalactic magnetic field
\cite{BO99}, it is unlikely to match the observed arrival direction
distribution. After M87, the next known nearby source is NGC315 which
is already too far at a distance of $\sim $ 80 Mpc. 

A recent proposal tries to get around
this challenge by invoking a  Galactic wind with a  strongly magnetized
azimuthal component \cite{ABMS99}. Such a wind can significantly alter
the paths of UHECRs such that the observed arrival directions of
events above 10$^{20}$ eV would trace back to the Virgo cluster close to 
M87. If our Galaxy has a such a wind is yet to be determined. The proposed
wind seems hard to support physically and would focus most events into the
northern Galactic pole and render point source identification
fruitless.\cite{BLS00} Future observations of UHECRs from the Southern
Hemisphere  by the Southern Auger Site will provide data on
previously unobserved parts of the sky  and help   distinguish plausible
proposals for the effect of local magnetic fields on arrival
directions. Full sky coverage is a key
discriminator of such proposals.   

{\it  AGN - Central Regions:}
The powerful engines that give rise to the observed jets and radio
lobes are located in the central regions of active galaxies and are
powered by the accretion of matter onto supermassive black holes. It
is reasonable to consider the central engines themselves as the likely
accelerators.\cite{T86,revmodels} In principle, the nuclei of  generic
active galaxies (not only the ones with hot spots) can accelerate
particles via a unipolar inductor not unlike the one operating in
pulsars. In the case of AGNs,   the magnetic field  is provided by the
infalling matter and the spinning black hole horizon provides the
imperfect conductor for the unipolar induction. 

The problem with AGNs as UHECR sources is two-fold: first, UHE particles
face  debilitating losses in the acceleration region due to the intense
radiation field present in AGNs,  and second, the
spatial distribution of objects should give rise to a  GZK cutoff of the
observed spectrum. In the central  regions of AGNs, loss processes are
expected to downgrade particle energies well below the maximum
achievable energy. This limitation has led to the proposal that  quasar
remnants, supermassive black holes in centers of inactive galaxies,  are
more effective UHECR accelerators.\cite{BG99} In this case, losses are
not as significant but the distribution of sources should still lead to a
clear GZK cutoff unless the spectrum is fairly hard.

{\it  Neutron Stars}
Another astrophysical system capable of accelerating
UHECRs is a neutron star.\cite{revmodels,Be92,BEO99}  Acceleration
processes inside the neutron star light cylinder are bound to fail
much like the AGN central region case:  ambient magnetic and radiation
fields induce significant losses. However, the plasma that expands
beyond the light cylinder is freer from the main loss processes and may
be accelerated to ultra high energies.

One possible source of UHECR past the GZK cutoff is the early evolution
of neutron stars. In particular, newly formed, rapidly rotating
neutron stars may accelerate iron nuclei  to UHEs  through relativistic
MHD winds beyond  their light cylinders.\cite{BEO99}  
In this case, UHECRs originate mostly in the
Galaxy and the arrival directions require that the primaries be  heavier
nuclei. Depending on the structure of   Galactic
magnetic fields, the trajectories of iron nuclei from Galactic
neutron stars may be consistent with the observed arrival directions of
the highest energy events.\cite{ZPPR98} Moreover,  if  cosmic rays 
of a few times $10^{18}$ eV are protons of Galactic origin, the
isotropic distribution observed at these energies is indicative of the
diffusive effect of the Galactic magnetic fields on iron at
$\sim 10^{20}$ eV. This proposal awaits a clear composition determination.

{\it  Gamma-Ray Bursts}
Transient high energy phenomena such as gamma-ray
bursts may accelerate protons to ultra-high energies .\cite{W95}
Aside from both having unknown origins, GRBs and UHECRs have some
similarities that argue for a common origin. Like UHECRs, GRBs are
distributed isotropically in the sky,  and
the average rate of $\gamma$-ray energy emitted by GRBs is comparable
to the energy generation rate of UHECRs of energy $>10^{19}$ eV in a
redshift independent cosmological distribution of sources, both have  $ 
\approx 10^{44}{\rm erg\ /Mpc}^{3}/{\rm yr} .$ 

However, the distribution of UHECR arrival directions and arrival
times argues against the GRB--UHECR common origin. Events past the GZK
cutoff require that only GRBs from $\la 50$ Mpc contribute. Since
less than about {\it one} burst is expected to have occurred within this
region over a period of 100 yr, the source would appear as a
concentration of UHECR events.   Therefore, a very large dispersion of  
$\ga$ 100  yr  in the arrival time of protons  produced in a single burst
is necessary. The deflection  by random magnetic fields combined with the
energy spread of the particles is usually invoked to reach the required
dispersion.\cite{W95} If the dispersion in time and space is
achieved, the energy spectrum for the nearby source(s) becomes very
narrowly  peaked  $\Delta E/E\sim1$. 
Finally, if the observed small scale clustering of arrival directions is
confirmed by future experiments with clusters having lower energy events
precede higher energy ones, bursts would be invalidated.\cite{SLO97}

\section{Hybrid Models}

The UHECR puzzle has inspired proposals that use Zevatrons to
generate UHE particles other than protons, nuclei, and photons.
These use physics beyond the standard model in a bottom-up approach,
thus, named hybrid models.

The most economical among such proposals involves a familiar
extension of the standard model, namely, neutrino masses.  If
some flavor of neutrinos have masses $\sim 1$ eV, the relic
neutrino background  will cluster in halos of galaxies and clusters of
galaxies. High energy neutrinos ($\sim 10^{21}$ eV) accelerated in
 Zevatrons  can annihilate on the neutrino background and
form UHECRs through the hadronic Z-boson decay.\cite{We97FMS97} 

This proposal is
aimed at generating UHECRs nearby (in the Galactic halo and Local Group
halos) while using Zevatrons that can be much further than the GZK
limited volume, since neutrinos do not suffer the GZK losses.The weak
link in this proposal is the nature of a Zevatron powerful enough to
accelerate protons above ZeVs that can produce ZeV neutrinos as
secondaries. This Zevatron is quite spectacular,
requiring an energy generation in excess of presently known highest
energy sources.

Another suggestion is that the UHECR primary is a new 
particle. The mass of a hypothetical
hadronic primary can be limited by the shower development of the Fly's
Eye highest energy event to be below $\la 50$ GeV.\cite{AFK98}  Both a
long lived new particle and the neutrino Z-pole proposals involve neutral
particles which are usually harder to accelerate (they are created as
secondaries of even higher energy charged primariess) but
can traverse large distances without being affected by the cosmic magnetic
fields. Thus, a signature of such hybrid models for future experiments is
a clear correlation between the position of powerful Zevatrons in the sky
such as distant compact radio quasars and the arrival direction of  UHE
events.\cite{FB98} 

Another exotic primary that can use a Zevatron to reach ultra high
energies is the vorton. Vortons are small loops of superconducting
cosmic string stabilized by the angular momentum of charge
carriers.\cite{DS89} Vortons can be a component of the dark matter in
galactic halos and be accelerated in astrophysical Zevatrons \cite{BP97}.
Although not yet clearly demonstrated, the shower development profile is
also the likely liability of this model.

\section {Top-Down Models}

It is possible that none of the astrophysical scenarios are able to 
meet the challenge posed by the UHECR data as more
observations are accumulated. In that case,  one alternative
is to consider top-down models.
This proposal dates back to the work on monopolonia of Hill and
Schramm.\cite{SH83HS83} Other top-down proposals involve 
the decay of ordinary and superconducting cosmic strings,  cosmic
necklaces, vortons, and superheavy long-lived relic particles. 
The idea behind these models is that relics of the very
early universe, topological defects (TDs) or superheavy relic (SHR)
particles,  produced  after or at the end of inflation, can
decay today and generate UHECRs.  Defects, such as cosmic strings,
domain walls, and magnetic monopoles,  can be generated through the
Kibble mechanism  as symmetries are broken with the
expansion and cooling of the universe.\cite{revmodels} 
Topologically stable defects can survive to the  present and
decompose into their constituent fields  as they collapse, 
annihilate, or reach critical current in the case of superconducting
cosmic strings. The decay products, superheavy gauge and higgs bosons,
decay into jets of hadrons, mostly pions.  Pions in the jets
subsequently decay into $\gamma$-rays, electrons, and neutrinos. Only a
few percent of the hadrons are expected to be nucleons.
Typical features of these scenarios are a predominant release of
$\gamma$-rays and neutrinos and a QCD
fragmentation spectrum which is  considerably harder than the case of
shock acceleration.

ZeV energies are not a challenge for top-down models since symmetry
breaking scales at the end of inflation typically are $\gg 10^{21}$
eV (typical X-particle masses vary between 
$\sim 10^{22} - 10^{25}$ eV) .  Fitting the observed flux
of UHECRs is the real challenge since the typical distances between TDs
is  the  Horizon scale,
$H_0^{-1} \simeq 3 h^{-1}$ Gpc. The low flux hurts proposals based on
ordinary  and superconducting cosmic strings. Monopoles usually suffer
the opposite problem, they would in general be too numerous. Inflation
succeeds in diluting the number density of monopoles   
usually making them too rare for UHECR production. To reach the
observed UHECR flux, monopole models usually involve some degree of
fine tuning. If enough monopoles and antimonopoles survive from the
early universe,  they may form a bound state, named monopolonium, that 
can  decay  generating UHECRs. The lifetime of monopolonia may be
too short for this scenario to succeed unless they are connected by
strings.\cite{PO99}

Once two symmetry breaking scales are invoked, a combination of
horizon scales gives room to reasonable number densities. This can be
arranged for cosmic strings that end in monopoles making a monopole
string network or even more clearly for cosmic necklaces.\cite{BV97}
Cosmic necklaces are hybrid defects where each monopole is connected to
two strings resembling beads on a cosmic string necklace. Necklace
networks may evolve to configurations that can fit the UHECR flux which
is ultimately generated by the annihilation of monopoles with
antimonopoles trapped in the string.\cite{BV97,BBV98} In these scenarios,
protons dominate the flux in the lower energy side of the GZK cutoff 
while photons tend to dominate at higher energies depending on the radio
background. If future data can settle the composition of UHECRs from 0.01
to 1 ZeV, these models can be well constrained.
In addition to fitting the UHECR flux, topological defect
models are constrained by limits on the flux of  high energy photons,
from 10 MeV to 100 GeV, observed by EGRET.

Another interesting possibility is the recent proposal that UHECRs are
produced by the decay of unstable superheavy relics that live much longer
than the age of the universe.\cite{BKV97}  SHRs may be produced at
the end of inflation by non-thermal effects such as a varying
gravitational field, parametric resonances during preheating,  instant
preheating, or the decay of topological defects. 
These models need to invoke special symmetries to insure unusually long
lifetimes for SHRs and that a sufficiently small percentage decays today
producing UHECRs.\cite{BKV97,CKR99KT99}  As in the topological defects
case, the decay of these relics also generate jets of hadrons.  These
particles behave like cold dark matter and could constitute a fair
fraction of the halo of our Galaxy. Therefore, their halo decay products
would not be limited by the GZK cutoff allowing for a large flux at UHEs.

Future experiments should be
able to probe these hypotheses.  For instance, in the case of SHR
and monopolonium decays, the arrival
direction distribution should be close to isotropic but show an
asymmetry due to the position of the Earth in the Galactic
Halo.\cite{BBV98} Studying plausible halo models and the expected
asymmetry will help constrain halo distributions especially when larger
data sets are available from future experiments. High energy gamma ray
experiments such as GLAST will also help constrain the SHR models due to
the products of the electromagnetic cascade.\cite{B99}

\section{Conclusion}

Next generation experiments  such as the High Resolution Fly's Eye which
recently started operating, the Pierre Auger Project which is now under
construction,  the proposed  Telescope Array, and
the OWL and Airwatch satellites  will 
significantly improve the data at the extremely-high end of the cosmic
ray spectrum.\cite{revdata} With these observatories a clear
determination of the spectrum and spatial distribution of UHECR
sources is within reach. 

The lack of a GZK cutoff should become clear with HiRes and Auger
and most extragalactic Zevatrons may be ruled out. 
 The observed spectrum will distinguish Zevatrons from
top-down models by testing power laws versus QCD fragmentation fits. 
The cosmography of sources should also become clear and able to
 discriminate  between plausible populations for UHECR sources. 
The  correlation of arrival directions  for events with
energies above
$10^{20}$ eV  with some known structure such as the Galaxy, the
Galactic halo, the Local Group or the Local Supercluster would be key
in differentiating between different models. For instance, a
correlation with the Galactic center and  disk should become apparent
at extremely high energies for the case of young neutron star winds,
while a correlation with the large scale galaxy distribution should
become clear for the case of quasar remnants. If SHRs or monopolonia are
responsible for UHECR production, the arrival directions should correlate
with the dark matter distribution and show the halo asymmetry. For these
signatures to be tested, full sky coverage is essential. Finally,  an
excellent discriminator would be an unambiguous composition
determination  of the primaries. In general, Galactic disk models  invoke
iron nuclei to be consistent with the isotropic distribution, 
extragalactic Zevatrons tend to favor proton primaries, while photon
primaries are more common for early universe relics. The hybrid detector
of the Auger Project should help determine the composition by measuring
simultaneously the depth of shower maximum and the muon content of the
same shower.  The
prospect of testing extremely high energy physics as well as solving
the UHECR mystery awaits improved observations that should be coming in
the next decade with experiments under construction or in the planning
stages.\cite{revdata} 

\medskip
 This work was supported by NSF through grant
AST 94-20759  and DOE grant DE-FG0291  ER40606.

\end{document}